\documentclass[final]{IEEEtran}
\usepackage[dvips,final]{graphicx}
\usepackage{latexsym}
\usepackage{amssymb}
\usepackage{amsfonts}
\usepackage{latexsym}
\usepackage{amsmath,epsfig,algorithmic,color,balance}
\usepackage{algorithm}
\usepackage{epstopdf}
\usepackage{subfig} % make it possible to include more than one captioned figure/table in a single float

\newcommand{\argmin}{\operatornamewithlimits{argmin}}
\newcommand{\argmax}{\operatornamewithlimits{argmax}}

\psfull

\title{Selection of network coding nodes for minimal playback delay in streaming overlays\thanks{This work has been partly supported by the Swiss National Science
Foundation, under grant PZ00P2-121906.}}

\author{Nicolae Cleju$^\ast$, Nikolaos Thomos$^\dagger$ and Pascal Frossard$^\dagger$\\
$^\dagger$Signal Processing Laboratory (LTS4),
Ecole Polytechnique F\'{e}d\'{e}rale de Lausanne (EPFL), Lausanne, Switzerland\\
$^\ast$The ``Gheorghe Asachi'' Technical University of Iasi, Iasi Romania.\\
nikcleju@etti.tuiasi.ro, \{nikolaos.thomos,pascal.frossard\}@epfl.ch}

\begin{document}

\maketitle

\begin{abstract}
Network coding permits to deploy distributed packet delivery algorithms that locally adapt to the network availability in media streaming applications. However, it may also increase delay and computational complexity if it is not implemented efficiently. We address here the effective placement of nodes that implement randomized network coding in overlay networks, so that the goodput is kept high while the delay for decoding stays small in streaming applications. We first estimate the decoding delay at each client, which depends on the innovative rate in the network. This estimation permits to identify the nodes that have to perform coding for a reduced decoding delay. We then propose two iterative algorithms for selecting the nodes that should perform network coding. The first algorithm relies on the knowledge of the full network statistics. The second algorithm uses only local network statistics at each node. Simulation results show that large performance gains can be achieved with the selection of only a few network coding nodes. Moreover, the second algorithm performs very closely to the central estimation strategy, which demonstrates that the network coding nodes can be selected efficiently in a distributed manner. Our scheme shows large gains in terms of achieved throughput, delay and video quality in realistic overlay networks when compared to methods that employ traditional streaming strategies as well as random network nodes selection algorithms.
\end{abstract}


\section{Introduction}
\label{sec:intro}

The recent development of overlay networks offers interesting perspectives for multimedia streaming applications, since network diversity can be used advantageously for improved quality of service. The traditional streaming systems based on ARQ or channel coding techniques however generally fail to efficiently exploit this diversity. They either suffer from relatively high computational cost, require coordination between network nodes or lead to suboptimal performance in large scale networks where local channel conditions are hard to estimate. A different paradigm has been initiated recently with network coding \cite{AhlswedeTIT00,LinearNC03}, where some processing is requested from the network nodes in order to improve the packet delivery performance. Specifically, network coding nodes combine buffered packets before forwarding them to next hop nodes. This coding strategy is particularly appealing in distributed streaming systems, as it removes the need for reconciliation between nodes. It locally adapts to the available bandwidth and packet loss rate and even permits to approach max-flow min-cut bound of the network graph. Overall, the network coding systems have shown improved resiliency to dynamics, delays, scalability and buffer capacities in networks with diversity \cite{WangR207}.

The application of network coding algorithms in multimedia streaming systems is however not straightforward. Specifically, multimedia streaming imposes strict timing constraints that impact the design of network coding algorithms. A practical network coding system has been presented in \cite{PracticalNC03} and addresses the specific characteristics of streaming applications. It implements randomized network coding (RNC) techniques \cite{RandomizedNC03} in the network nodes and devises a protocol to deal with buffering issues and timing constraints. Moreover, it introduces the concept of generations that restricts coding operations to packets that share similar decoding deadlines. However, network coding systems still face important issues in practical systems due to the decoding delays imposed by successive network coding operations. This delay as well as the computational overhead in the system grow with the number of network coding nodes. It becomes therefore important to select efficiently the subset of nodes that perform network coding in order to control delay and complexity and still exploit efficiently the diversity in the network.

In this paper, we discuss solutions for the selective placement of a few network coding nodes in order to reduce the delay for video delivery. The nodes in the network are categorized into {\emph{network coding}} (NC) and {\emph{store and forward}} (SF) nodes. The network coding nodes use the practical network coding algorithm described in \cite{PracticalNC03}, which has been selected for its effectiveness and simplicity. Similarly, we adopt the concept of coding generation and buffer models \cite{PracticalNC03} for proper handling of the timing constraints in the stream delivery. We first build on our previous work \cite{ThomosICC10} and estimate the rate of non-redundant packets in each network node. This rate is an indication of the goodput of the system as it measures the number of useful and non-redundant packets that are received at a node. It permits to estimate the decoding delay in the client nodes; this corresponds to the time necessary for collecting enough useful packets to build a full rank decoding system. We use the delay estimation to select the subset of nodes that should implement network coding such that a maximal goodput or a minimal delay is attained. We propose two algorithms that iteratively choose the SF nodes to be turned into NC nodes for improving the system performance. Both algorithms differ in their view of the network status. The first algorithm assumes that a central node has complete knowledge about the status of the overlay network in terms of available bandwidth and packet loss rate. The second algorithm only uses local estimations of the network status at each node and provides a solution for distributed systems. The simulation results show that the proper selection of only a few network coding nodes already leads to throughput gains that come close to max-flow min-cut bound and greatly decrease the delay necessary for media stream delivery. Moreover, the algorithm that only considers local network statistics performs very competitively with the algorithm that uses full knowledge of the network topology. Both algorithms even select the same nodes for network coding in most of the cases. Furthermore, they both outperform basic streaming algorithms built on store and forward approaches as well as  solutions where network coding nodes are selected randomly. These observations are confirmed in realistic overlay networks where our method can improve users' video experience even in the case where only few nodes implement randomized network coding. This is due to a good balance between decoding delay and efficient use of the network diversity. Finally, minimal network knowledge is often sufficient for determining the efficient positioning of the network coding nodes. 

The paper is organized as follows. In Section \ref{sec:framework}, we present the framework under consideration and briefly overview the network coding principles. In Section \ref{sec:buffer}, we describe the model of the SF node buffer that is eventually used for delay computation. Then we present in Section \ref{sec:innovativeflows} a methodology for estimating the useful flow rate in the network nodes as well as the decoding delay. The centralized and semi-distributed algorithms for selecting the network coding nodes are presented in Section \ref{sec:algos}. Simulation results are proposed in Section \ref{sec:res} where the benefits of the proposed algorithms are evaluated for video streaming applications in various realistic network cases. Finally, the related work is discussed in Section \ref{sec:related} and conclusions are drawn in Section \ref{sec:conclusions}.

\section{Network Coding Framework}
\label{sec:framework}

%\subsection{System description}

\begin{figure}[t]
\begin{center}
~\includegraphics[width=0.9\columnwidth]{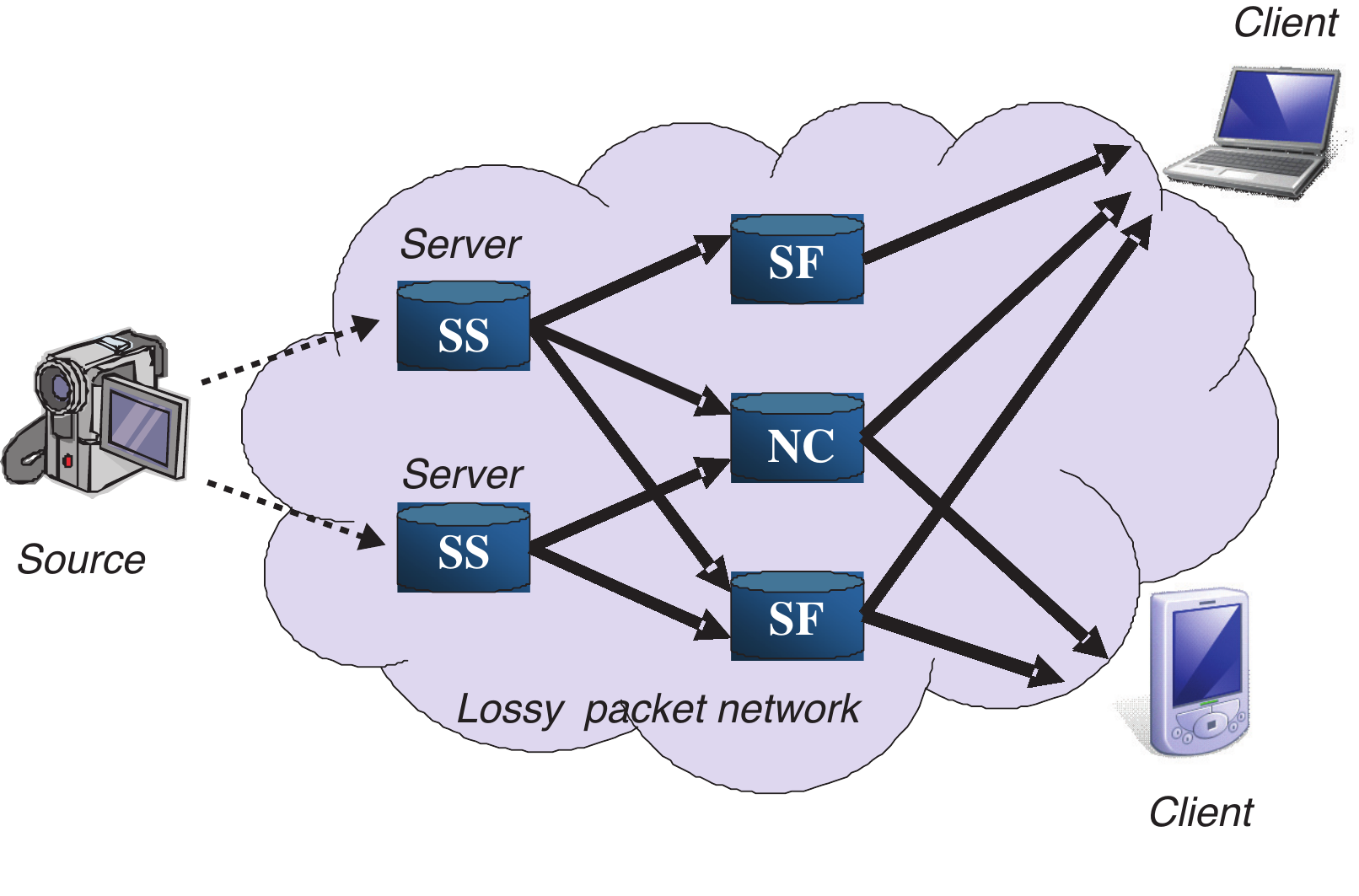}~\
\end{center}
\caption{Illustration of a system for streaming on overlay
networks. Multiple streaming servers (SS) send information to
clients on a lossy packet network via intermediate nodes that can
be either network coding (NC) or 'store and forward' (SF) peers.}
\label{fig:framework}
\end{figure}

We consider a streaming system that consists of servers, clients and intermediate nodes, as illustrated in Fig. \ref{fig:framework}. The overlay network offers source and path diversity, which can be efficiently exploited with network coding techniques that randomly combine packets in the nodes. This increases the packet diversity in the network and leads to efficient exploitation of the channel resources without the need for complex scheduling or nodes coordination mechanisms  \cite{AhlswedeTIT00}. The network is modeled by a directed acyclic graph $G=(V,E)$ where $V$ is the set of network nodes and $E$ is the set of edges (links) in the network. Each network link between nodes $u$ and $v$ is characterized by a bandwidth $b_{u,v}$ (expressed in terms of packets per second) as well as a packet loss rate $\pi_{u,v}$. We assume that all servers transmit the same multimedia content to clients via intermediate nodes that could either be network coding (NC) or ``store and forward'' (SF) nodes. We consider that the intermediate nodes are not necessarily interested in the transmitted content, but rather act as helper nodes and assist the packet delivery system. The system implements a push-based packet delivery strategy that involves lower communication and coordination overhead than a pull-based solution. We consider that the servers can also implement randomized coding on the source packets for improved robustness. The coded packets are then pushed to the clients through the successive intermediate nodes. Finally, the clients perform network decoding after receiving enough packets to build a full rank decodable system of packets.

The SF nodes simply send at each transmission opportunity the first packet in their buffer, which has not been sent previously. The buffer is managed in a first-in-first-out manner, where the oldest packets are replaced by new ones when the buffer is full. When the outgoing bandwidth is larger than the incoming capacity, a SF node sends random replicates of packets from its buffer. On the other hand, the intermediate nodes that perform network coding combine randomly the buffered packets in order to generate network coded packets that are further transmitted to neighbor nodes. As suggested in \cite{PracticalNC03}, the NC nodes first check whether the received packets are {\emph{innovative}}, where innovative packets characterize packets carrying novel information. Non-innovative packets are discarded immediately as they do not increase the symbol diversity into the network. Then the NC nodes randomly combine the remaining packets with coding operations based on randomized network coding (RNC) \cite{RandomizedNCTITOct06}.  It is a simple and efficient network coding solution in distributed systems. RNC codes work similarly to rateless codes \cite{Shokrollahi06,Luby02} and can generate an arbitrary  number of coded packets from a given set of source packets. It provides a means for simple bandwidth adaptation.

\begin{figure}[t]
\begin{center}
\includegraphics[width=7.5cm]{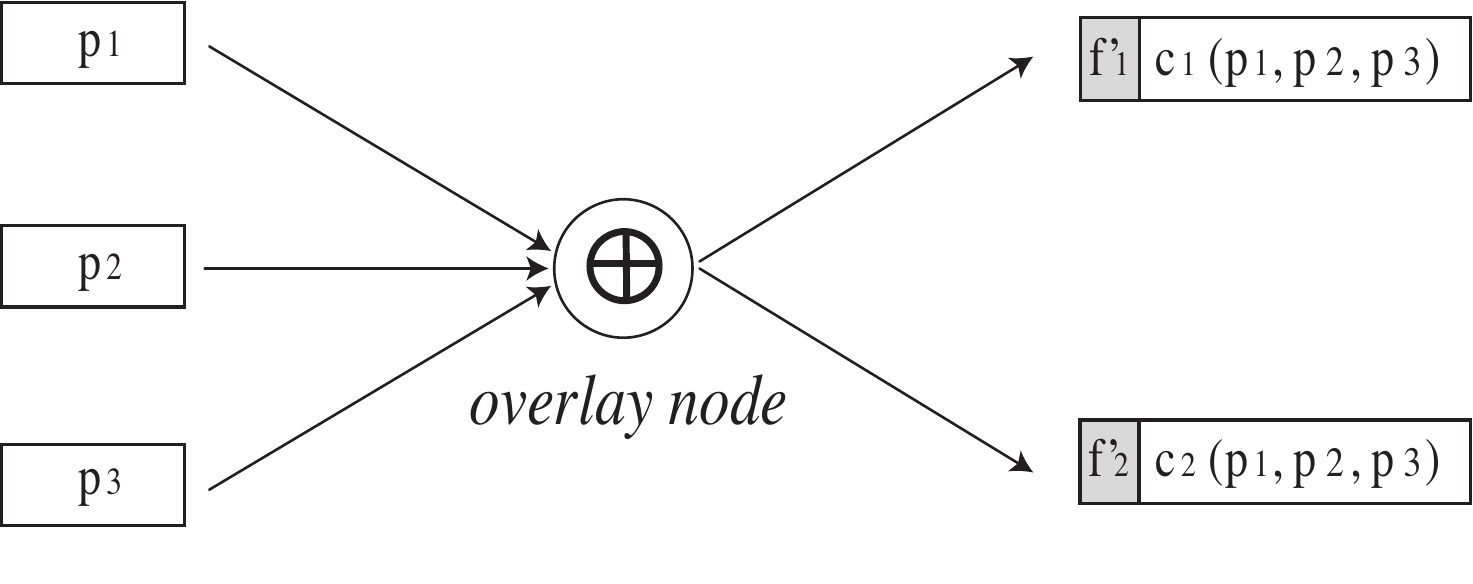}
\end{center}
\caption{A NC node combines incoming packets $p_i$ and generates network coding packets $c_m$. A header $f'_m$ is appended to each coded packet and carries the coding coefficients.} \label{fig:ncpeer}
\end{figure}

Formally, the network coding operations performed in a peer node can be written as follows. A NC node $u$ generates $M$ packets by RNC. The $m$\textsuperscript{th} network coded packet $c_m$ is of the form
\[ c_m=\sum\limits_{p_i(u) \in \mathcal{N}(u)} f_{m,i} \cdot p_i(u)
\]
where $\mathcal{N}(u)$ corresponds to the set of packets of the same generation that are available at node $u$, $p_i(u)$ denotes either a network coded packet or a native (uncoded) packet, and $f_{m,i}$ is a random coefficient over the Galois field of size $q$, GF(q). The basis of the Galois field is typically set to $q=256$, as it has been shown in \cite{PracticalNC03} that this guarantees high symbol diversity and low probability of building duplicate packets. As the packets combined in a node are actually combinations of the original data packets, the encoded packets can be expressed as a function of the native packets
\begin{equation}
c_m=\sum\limits_{i=1}^{N} f_{m,i}^{'} \cdot n_i \label{eq:coeff}
\end{equation}
where $N$ is the total number of native packets, e.g., for video transmission $N$ can be the number of video packets. The parameters $n_i$ and $f_{m,i}^{'}$ represent respectively the native packets and their corresponding coding coefficients after random network coding operations. It is worth noting that some of the coefficients $f_{m,i}^{'}$ can be zero, which means that $c_m$ does not contain information about the native packet $n_i$. As the coding coefficients are chosen randomly, a header of constant length is appended to each packet with coefficient information, so that the received can decode the stream and recover the original data packets. A network coded packet is thus augmented with a header containing the vector of coding coefficients. Fortunately, the header does not grow with the number of hop transmissions due to Eq. (\ref{eq:coeff}). The encoding procedure in a peer node is depicted in Fig. \ref{fig:ncpeer}.

The decoding operations at the client basically consist in solving the system of equations that correspond to the network coding operations. Upon collecting a network coded packet, the client stores it in a buffer and adds a line into a matrix $\mathbf{F}$ that contains the coding coefficients. When a full rank system is collected, the original packets are reconstructed by solving the following equations
\[
\mathbf{c}=\mathbf{F} \cdot \mathbf{n} \Longrightarrow \left[
{\begin{array}{*{20}c}
   {c_1 }  \\
    \vdots   \\
   {c_N }  \\
\end{array}} \right] = \left[ {\begin{array}{*{20}c}
   {f_{1,1}^{'} } &  \cdots  & {f_{1,N}^{'} }  \\
    \vdots  &  \ddots  &  \vdots   \\
   {f_{N,1}^{'} } &  \cdots  & {f_{N,N}^{'} }  \\
\end{array}} \right] \cdot \left[ {\begin{array}{*{20}c}
   {n_1 }  \\
    \vdots   \\
   {n_N }  \\
\end{array}} \right]
\]
where $\mathbf{c}$ and $\mathbf{n}$ are respectively vectors with the coded and source packets. The solution of the equations system is typically computed by gaussian elimination \cite{PracticalNC03}. 

The proposed streaming system leads to the following observations. First, the network coding nodes act somehow similarly to sources in the sense that they refresh the set of packets in the network by coding operations. This is necessary as there is a non-zero probability for the reception of duplicate packets in the network nodes when the network is mostly composed of SF nodes. These duplicates can be generated by a node that does not receive enough diverse packets, or from different nodes that independently transmit identical packets. These duplicate packets lower significantly the stream delivery performance especially in networks containing bottlenecks. However, the careful placement of a few network coding nodes in the overlay can help to reduce the number of duplicates in the network. If the number of network coding nodes however becomes too large, the probability for the randomized network coding operations to generate duplicate packets becomes again non-negligible. This is especially the case if coding operations are restricted to small generations due to delay constraints. As redundancy, delay and computational overhead might increase with the number of coding nodes, it becomes quite apparent that efficient systems should not implement network coding in every overlay node. Instead, one has to find an effective placement of network coding nodes in order to fully exploit the network diversity in overlay streaming applications.

%===============================================================
\section{SF buffer model}
\label{sec:buffer}
%===============================================================

\begin{figure}
  \centering
	\includegraphics[scale=0.33]{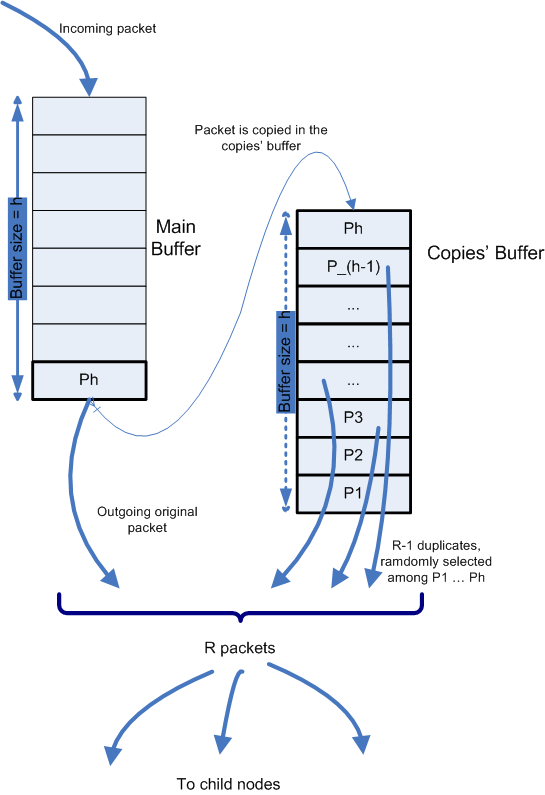}
	\caption{Packet replications procedure followed in a SF node after the reception of the $h^{th}$ innovative packet.}
	\label{fig:fig_bufferPh}
\end{figure}

We provide in this section a buffer model for the SF nodes, which forward and possibly replicate packets if the outgoing bandwidth is sufficient. The buffer model is used to estimate the rate of replicated packets, which is an important parameter in the computation of the decoding delay in the receivers.

As illustrated in Fig.  \ref{fig:fig_bufferPh}, we consider that each SF node has two buffers of capacity $h$ (in packets): the Main Buffer (MB), where the incoming packets are stored, and the Copies Buffer (CB) where copies of the packets that have been recently transmitted are stored. Both buffers follow a FIFO model as the oldest packets are overwritten by the new ones when the node's buffer capacity is exceeded. In addition, since our system works with deadline-constrained data, the packets are removed from the buffers when their decoding deadline expires. 

The buffering process works as follows: when a packet arrives in a SF node, it is stored in MB. When the SF node has a transmission opportunity, a packet from MB is sent and thus removed from MB. A copy of this packet is kept in CB. Whenever MB is empty and the node has other transmission opportunities, it randomly selects a packet from CB and transmits it. In other words, the node transmits packet replicates when the outgoing bandwidth is sufficient. The packets in CB are overwritten after some time by newer packets. In our model, if a SF node does not have sufficient outgoing bandwidth for replication, it does not use CB. Alternatively, if the outgoing bandwidth is large, CB is used extensively and MB is often empty. 

We are now interested in computing the number of packet replicates generated by the SF node $u$ under the proposed buffer model. A priori, the average number of replicates per packet $R(u)$ at node $u$ is given by the ratio of outgoing and incoming bandwidths $b_o(u)$ and $b_i(u)$ respectively. We can write $R(u) = \frac{b_o(u)}{b_i(u)}$. However, the probability for a packet to be replicated depends on the order of its arrival to the SF node. Typically, a packet that arrives early spends more time in the buffer and thus has a higher probability to be replicated than a packet that arrives late and close to the decoding deadline. We thus consider three cases depending on the order of arrival of the packets. We denote the arrival by the position $k$ of a packet in a coding generation. The first case includes the earliest packets, which reach the node while CB is not full. Every packet is replicated with a uniform probability in CB, but the packets that stay longer in the buffer have a higher chance to be replicated. The number of copies for the $k^{th}$ packet is thus given by:

\begin{equation} \displaystyle
R_k(u) = 1 + (R(u)-1) \left( \sum_{x=k}^{h}{\frac{1}{x}} + (k-1)\frac{1}{h}\right), \ \mbox{for} \ k \in [1,h]
\label{eq:CPk}
\end{equation}

The second case corresponds to packets that reach the node while CB is full. When CB is full, each new packet overwrites the oldest packet in CB. Each packet has a lifetime in CB that corresponds to the time necessary to collect $h$ new packets in the SF node. The replication probability is equal to $1/h$ and the number of copies is then equivalent to the average replication rate in the node. In this stationary mode, we have 

\begin{equation} \displaystyle
R_k(u) = R(u) , \quad \mbox{for} \ k \in [h+1,K] , 
\label{eq:CPk2}
\end{equation}
where $K$ is the number of packets that fully traverse CB until the head of the buffer's queue. Finally, the third case corresponds to the packets that do not spend a full lifecycle in CB due to the expiration of the decoding deadline. When the decoding deadline expires, CB is flushed, and the packets in the buffer at that moment have less opportunities to be replicated, since they do not traverse fully the buffer. If $k' = k - K$ denotes the position of the packets in CB when the buffer is flushed, the number of copies of the late packets can then be written as 

\begin{equation} \displaystyle
R_k(u) = 1 + (R(u)-1) \cdot \frac{ h - k' + 1 }{h}, \ \mbox{for} \ k \in  [K+1, |\mathcal{N}(u)|] ,
\label{eq:CPosi}
\end{equation}
where $|\mathcal{N}(u)|$ is the number of packets of the same coding generation that reach the node $u$.

In summary, two main factors affect the packet replication rate, the FIFO behavior of the CB buffer and the expiration of the decoding deadline that causes the deletion of packets in the buffer. Thus, the first $h-1$ packets are replicated more than average and the last packets are replicated less than average, while the intermediate packets  have constant replication rate. Finally, it should be noted that, depending on the bandwidth value and the delay constraints, there are situations where the buffer does not reach the stationary regime of Eq. (\ref{eq:CPk2}) and the computation of the number of replicates shall be adapted accordingly. 

We use the above buffering model to compute an equivalent packet replication rate $\hat{R}(u)$ for all packets in a SF node, which is more precise than the average value $R(u) = \frac{b_o(u)}{b_i(u)}$. The equivalent replication rate is estimated so that the number of packets at the client $c$ is preserved with respect to the case where the packet replication rate is computed independently for each packet. We assume that each packet travels independently to the client $c$, and we pose the following equivalent condition:

\begin{equation} \displaystyle
\sum_{k=1}^{|\mathcal{N}(u)|} \left\{1 - \left(\epsilon_c(u) \right)^{R_k(u)}\right\}  =  |\mathcal{N}(u)| \cdot  \left\{1 - \left(\epsilon_c(u)\right)^{\hat{R}(u)}\right\} ,
\label{eq:equivRech}
\end{equation}
where $\epsilon_c(u)$ is the probability of loosing a packet between the node $u$ and the client $c$. The number of copies $R_k(u)$ is computed from Eqs. (\ref{eq:CPk}), (\ref{eq:CPk2}) and (\ref{eq:CPosi}). Rewriting the above equation, we can express the equivalent replication rate as

\begin{equation} \displaystyle
\hat{R}(u) = \frac{\log{\left(1 - \frac{\sum_{k=1}^{|\mathcal{N}(u)|} \left\{1 - \left(\epsilon_c(u)\right)^{R_k(u)}\right\}}{|\mathcal{N}(u)|}\right)}}{\log{\left(\epsilon_c(u)\right)}} .
\label{eq:Rech}
\end{equation}

We use this replication rate estimate in the computation of the decoding delay in the next section.

%*****************************************************************************************************
%*****************************************************************************************************
\section{Delay analysis}
\label{sec:innovativeflows}
%*****************************************************************************************************
%*****************************************************************************************************

%*****************************************************************************************************
\subsection{Estimation methodology}
%*****************************************************************************************************

\begin{figure*}
  \centering
  %\subfloat[From the source.]
  \subfloat[]
  {
		\includegraphics[scale=0.2]{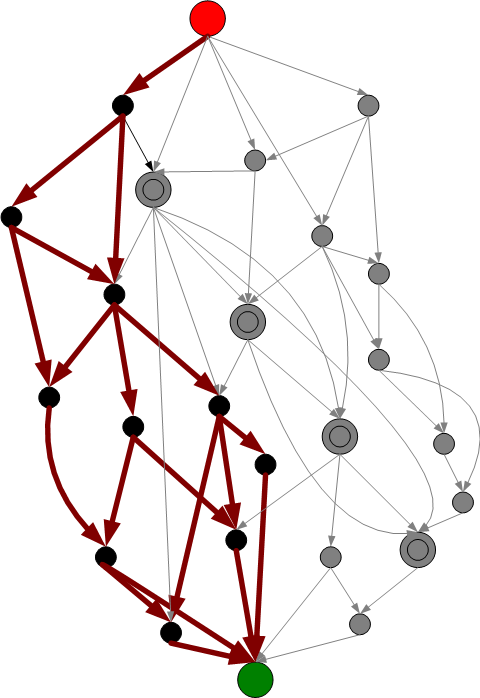}
		\label{fig:networkflows:source}
	}
  %\subfloat[\centering From the first \\NC node.]
  \subfloat[]
  {
		\includegraphics[scale=0.2]{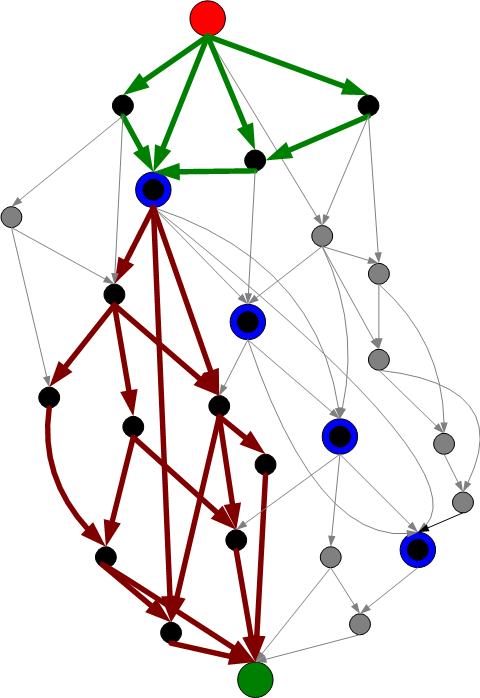}
		\label{fig:networkflows:NC1}
	}
  %\subfloat[\centering From the second \\NC node.]
  \subfloat[]
  {
		\includegraphics[scale=0.2]{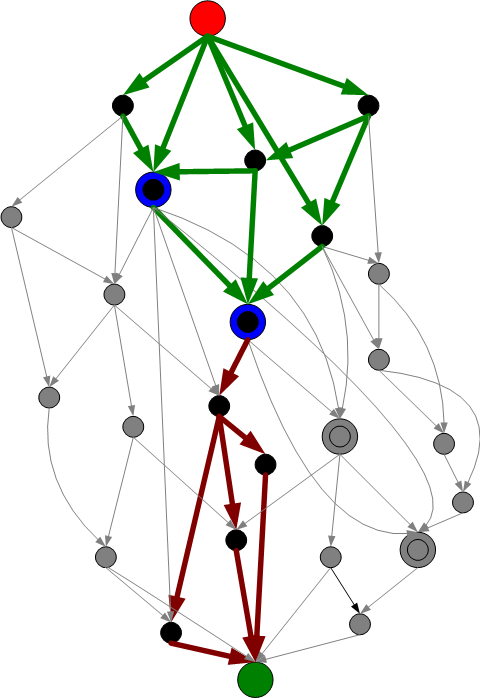}
		\label{fig:networkflows:NC2}
	}
  %\subfloat[\centering From the third \\NC node.]
  \subfloat[]
  {
		\includegraphics[scale=0.2]{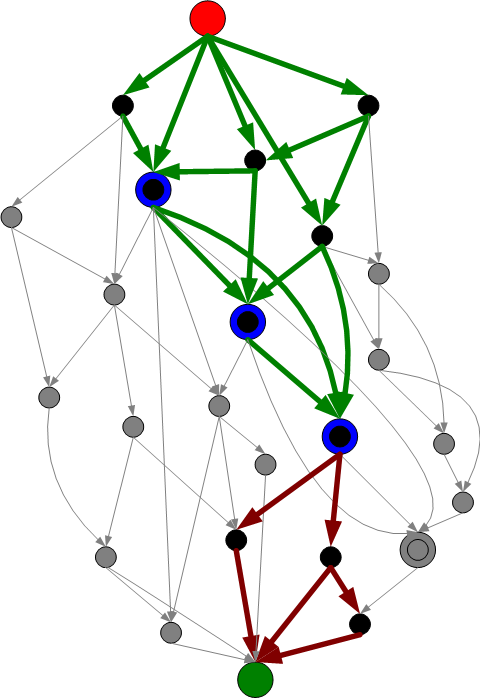}
		\label{fig:networkflows:NC3}
	}
  %\subfloat[\centering From the fourth \\NC node.]
  \subfloat[]
  {
		\includegraphics[scale=0.2]{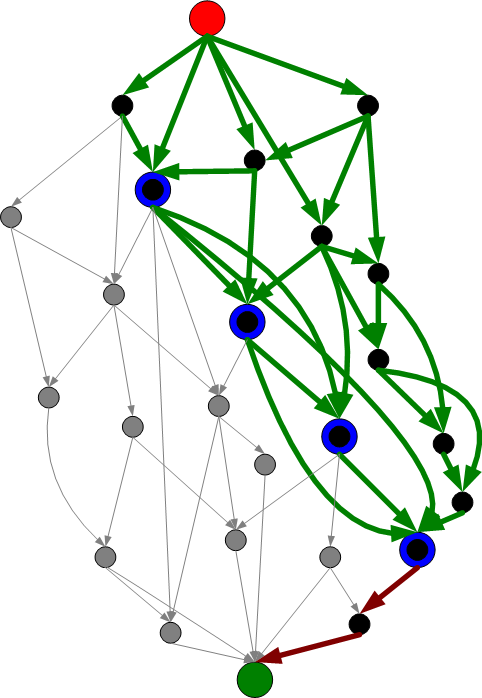}
		\label{fig:networkflows:NC4}
	}
	\caption{Flow decomposition of the network graph by the proposed algorithm. The top (red) node and the bottom (green) node are respectively the source and the client of the considered topology. The SF and NC nodes are represented respectively by the big and small black nodes. Flow from the (a) source, (b) first NC node, (c) second NC node, (d) third NC node, and (e) fourth NC node.}
	\label{fig:networkflows}
\end{figure*}

Our objective is to minimize the decoding delay by the proper placement of NC nodes in the overlay. The decoding delay is the time required to gather a sufficient number of innovative packets for decoding. In the analysis below we restrict our attention to cases where a full rank system is built at decoder and estimate the delay necessary for this situation to happen. We further construct our analysis for one coding generation, while the extension to multiple generations is straightforward. The decoding delay depends on the rate of innovative packets at the client. The innovative rate increases monotonically with the number of useful packets at the client \cite{PracticalNC03}, which corresponds to the number of different packets that reach the client. Hence, a higher rate of useful packets leads to a smaller decoding delay. 

In order to compute the decoding delay, we consider that SF nodes replicate packets in case of large outgoing bandwidth. We further consider that new packets are generated only at sources and NC nodes. We treat these nodes independently in the computation of the delay at the client as illustrated in Fig. \ref{fig:networkflows}. We assume that the probability of generating two identical packets in the sources or NC nodes is negligible due to the large size of the Galois field. In more details, we first estimate the delay noticed by the client when packets are sent from a given source through all the paths connecting this source to the client, except for the paths that traverse NC nodes (see Fig. \ref{fig:networkflows:source}). Next, we consider the NC node that is the closest to the source and all the paths that connect it to the client, except for those passing through other NC nodes (see Fig. \ref{fig:networkflows:NC1}). Similarly, all other NC nodes are considered only when all their parent NC nodes have been visited. This procedure is repeated for the unprocessed NC nodes and the corresponding graphs are shown respectively in Figs. \ref{fig:networkflows:NC2}, \ref{fig:networkflows:NC3} and \ref{fig:networkflows:NC4}. The total delay is computed under the assumption that all sources and NC nodes send independent streams. Equivalently, we assume that the total useful flow is equal to the sum of the useful flows generated by the source and all the NC nodes.

As we consider lossy network paths, we have to consider that some of the packets generated by network nodes do not reach their destination. We thus estimate the probability $\epsilon_c(u)$ that a packet sent by the node $u$ does not reach the client $c$. This probability is computed on the subgraph that contains all the paths connecting the node $u$ with the client $c$ but excludes the paths that traverse other NC nodes since these typically alter the set of received packets. The corresponding subgraphs are colored in red in Fig. \ref{fig:networkflows}. We denote by $D(u)$ the set of children of node $u$ (excluding the NC nodes) in the subgraph and we define \[\rho_{u,v} = \frac{\displaystyle b_{u,v}}{\displaystyle \sum_{v \in D(u)}b_{u,v}}\] as the probability that a packet transmitted by node $u$ is forwarded to a descendant node $v \in D(u)$. In addition, we write the probability that a packet is deleted in a node due to buffer overflow as \[
\beta(u) = \begin{cases}
        1- \frac{b_o(u)}{b_i(u)}, & b_o(u) < b_i(u) \\
        0, & b_o(u) \ge b_i(u)
        \end{cases}
\]
Recall that $b_o(u)$ and $b_i(u)$ are respectively the cumulative incoming and outgoing bandwidth of node $u$. Then, a packet sent by node $u$ might not reach the client $c$ due to one of the following three causes. The packet can be lost during its transmission to the child node $v$ or it can be lost at the node $v$ due to buffer overflow. Finally, it can be lost along with all its possible copies during the transmission from the child node $v$ to the client $c$. Overall, the probability $\epsilon_c(u)$ is given  as 

\begin{equation}
\begin{split}
\epsilon_c(u) = & \sum_{v \in D(u)}  \rho_{u, v} \cdot \left\{ \pi_{u,v} + 
 \left( 1-\pi_{u,v} \right) \cdot \beta(u) \right\}
\\ + & \sum_{v \in D(u)}  \rho_{u, v}  \left( 1-\pi_{u,v} \right) \cdot (1- \beta(u)) \cdot \epsilon_c(v)^{\hat{R}(v)} .
\end{split}
\label{eq:p0}
\end{equation}

The probability $\epsilon_c(u)$ can be computed recursively backwards starting from the clients up to the server or NC nodes. Specifically, we first set to zero the loss probabilities for all the clients. Then, all nodes in the directed acyclic subgraph are visited backwards. Each node is visited only when all its children nodes have been processed already. Once $\epsilon_c(u)$ is known, we can compute the rate of useful packets received by the client $c$ from a source or a NC node by multiplying the respective outgoing rate by the probability of correct reception $1 - \epsilon_c(u)$. 

Next, we define $N_c(u)$, the number of packets received by any node $u$ that are potentially useful for the client $c$. These packets correspond to the data that reaches the node $u$ and contains information that is potentially useful for the client $c$. It depends on the paths connecting the source nodes and the NC nodes to the node $u$, \textit{i.e.}, the subgraphs colored in green in Fig. \ref{fig:networkflows}. It also depends on the set of SF nodes on these paths. The rate of useful packets transmitted by the sources corresponds to their outgoing bandwidth, as they are able to generate any  number of different packets via network coding. However, the useful rate in NC nodes may be smaller than their outgoing bandwidth, as they may have only part of the source information in their buffer. Finally the number of useful packets in SF nodes cannot be larger than the number of incoming packets. It is however difficult to estimate in a direct way, so that the estimation of the delay based on the useful rate sent by network nodes becomes hard. We therefore propose below to compute the delay in a recursive manner.

%===============================================================
\subsection{Decoding delay}
%===============================================================

In this section, we estimate the delay $t_c$ at a client node $c$. The decoding delay depends on the rate of useful packets received from the multiple sources or network coding nodes. We estimate the time necessary to form a system of full rank $G$ at the decoder, where $G$ is the generation size in packets. In practice, the client might need to collect a slightly larger number of packets, $\tilde{G} = G/(1 - x)$ \cite{PracticalNC03} for forming a system with $G$ innovative packets. This is due to the possibility that useful packets from a source might still be redundant and not completely independent of packets generated by other sources. The exact value of the overhead factor $x$ depends on the coding system (e.g., it can be upper-bounded by $1/q$ for RNC, where $q$ is the  GF size). However, our analysis is relative, and compare different configurations to select the option that leads to the minimal decoding delay. It becomes therefore equivalent to work with $G$ or $\tilde{G}$ since the solutions that lead to the fastest delivery of $G$ and respectively $\tilde{G}$ packets are identical. We choose to work with $G$ in the rest of this section. 

We compute the average decoding delay by first estimating the time necessary to collect enough packets from each source or NC node independently. Under the assumption that each one of these multiple collection processes represents a uniform flow of packets, we can finally approximate the expected decoding delay as the time necessary for the collection of a sufficient number of packets from multiple independent flows. 

The complete algorithm for computing the decoding delays is given in Algorithm \ref{algo:algo_innovative_refinedrates}. Note that the algorithm uses an iterative procedure to compute the decoding delays, since the equivalent packet replication rate in SF node (see Section \ref{sec:buffer}) cannot be exactly computed at first. The algorithm initializes the replication rate to an average value given by the input and output bandwidths of each node, and refines this value along with the successive decoding delay estimations. The NC nodes are examined in the order of their proximity to the sources, \textit{i.e.}, the nodes that are closer to the sources are processed first. The number of useful packets $N_c(u)$ is computed recursively at all NC nodes, starting from those that are close to the sources. Then the algorithm considers NC nodes that receive packets from NC nodes that have been already visited. This specific procedure is applicable in our framework as we consider the iterative selection of network coding nodes. Nodes are checked in a greedy way and the algorithm improves at each stage the current solution by the selection of the additional network coding node that brings the largest delay reduction. The overall algorithm typically converges only after a few iterations. 

We describe now the delay estimation algorithm in more details. The steps 7-11 of Algorithm \ref{algo:algo_innovative_refinedrates} correspond to the estimation of the useful rate $N_c(u)$ in NC nodes, which is necessary for computing the initial replication rate of the node $u$. As this rate is difficult to estimate in a direct way, we choose a differential method by comparing the delay $t_c(u)$ observed by the client when the node $u$ is active and respectively silent. From the delay difference $\Delta t_c(u)$ we compute the rate difference $\Delta N_c(u)$ that is the useful rate at node $u$. Finally, the rate $N_c(u)$ of the packets at node $u$ that are useful for the client $c$ is computed by solving
\begin{equation} \displaystyle
\Delta N_c(u) = \begin{cases}
  N_c(u) \cdot \left( 1 - \epsilon_c(u)^{\hat{R}(u)} \right),  &  b_o(u) > b_i(u) \\
  N_c(u) \cdot \frac{b_o(u)}{b_i(u)} \cdot \left( 1 - \epsilon_c(u) \right),  &  b_o(u) < b_i(u) 
  \end{cases}
\label{eq:D_cn} 
\end{equation}
where the first and second conditions correspond to the cases when the node $u$ in SF mode has a small incoming, respectively outgoing bandwidth. Note that, when the network does not contain any NC nodes, the rate of useful packets that are transmitted is simply equal to the output bandwidth of the sources. In this case, the useful rate received by the client is simply $N_c(s) = b_o(s) \cdot (1-\epsilon_c(s))$. 

Then we compute the delay due to packets sent by the different sources and NC nodes in the network. We consider two cases. First, we consider the NC nodes that have limited incoming bandwidth (i.e., $N_c(u) < b_o (u)$ in line 12 or Alg. 1) and the sources with outgoing bandwidth larger than the source rate. The probability of generating useful packets in such nodes evolves as the buffer fills in. We start by estimating the number of different packets received by the client $c$ when the node $u$ is the only source of information. In average, the node $u$ sends  $\nu(u) = \frac{b_o(u)}{N_c(u)}$ packets in the inter-arrival time of two consecutive packets in its buffer, which are combinations of the same set of input packets. Out of these $\nu(u)$ packets, $k$ packets can be considered as useful for the client $c$ if the decoding system has a rank deficiency of $k<\nu(u)$. Furthermore, due to packet losses and bandwidth variations in the network, each of the packets generated by the node $u$ arrives at the client $c$ with probability $\epsilon_c(u)$. The probability $A_k(u)$ that $k$ out of the $\nu(u)$ packets arrive at the client $c$ is 
\begin{equation} \displaystyle
      A_k(u) = {{\nu(u)}\choose{k}} \left( 1 - \epsilon_c(u)\right)^k \epsilon_c(u)^{\nu(u)-k}
\label{eq:pk}
\end{equation}
Note that in general, $\nu(u)$ has not an integer value. We therefore perform an interpolation between the values of $A_k(u)$ evaluated on the integer values nearest to $\nu(u)$.

We then consider the probability $P_c(u,r,n)$ for the client $c$ to collect $r$ useful packets from data sent by a node $u$ that possesses $n$ useful packets. This probability can be computed recursively as

\begin{equation} 
        P_c(u,r,n) = \sum_{k=0}^{\nu(u)} P_c(u,r-k,n-1)  \cdot  A_k(u), \forall r \in [1.. n-1]
\label{eq:Pr<N}
\end{equation}

Eq. (\ref{eq:Pr<N}) holds for $r<n$. When $r=n$, it becomes
\begin{equation} 
        P_c(u,n,n) = \sum_{k=1}^{\nu(u)} {P_c(u,n-k,n-1)  \cdot  \sum_{l=k}^{\nu(u)} A_l(u)}
        \label{eq:Pr<N2}
\end{equation} 

We further denote by  $\mathcal{P}_c(u,r,n)$ the probability that the client $c$ collects $k$ useful packets precisely due to the arrival of the $n^{th}$ useful packet in the sending node $u$. It can be written as

\begin{equation} 
        \mathcal{P}_c(u,r,n) = \sum_{k=1}^{\nu(u)} \left[ P_c(u,r-k,n-1)  \cdot  \sum_{l = k}^{\nu(u)} A_l(u) \right],
\label{eq:Pr<Nbis}
\end{equation}
where $P_c(u,r-k,n-1)$ includes all possible events that lead the node $u$ to collect packets with rank $r-k$ when it receives the $n-1$ useful packet for client $c$.

The arrival time of the $n^{th}$ useful packet at the sending node $u$ can be computed from the useful packet rate $N_c (u)$.% following Algorithm \ref{algo:algo_Nc}. 
We assume that $N_c(u)$ represents a  constant rate and that the arrival times of packets in node $u$ are uniformly distributed. Now, one can compute the expected number of useful packets $E_c (u)$ that are necessary at the sending node $u$ for the client $c$ to receive $G$ useful packets. It is expressed simply as 

\begin{equation} \displaystyle
        E_c (u) = \sum_{p=G}^\infty {p \cdot \mathcal{P}_c(u,G,p)}
\label{eq:Nbar}
\end{equation}

The decoding delay for the client $c$ when the NC node $u$ is the only source of information can be estimated by dividing the expected number of necessary packets by the inter-arrival time between two useful packets. It is written as

\begin{equation} \displaystyle
        t_c (u) = {E_c (u)}/{N_c (u)}
\label{eq:tbar}
\end{equation}

Then, we consider the sources and the NC nodes that are over-provisioned in bandwidth (i.e., the set of nodes $u$ where $N_c(u) > b_o (u)$, line 14 in Alg. 1). We assume that they transmit packets that are all potentially useful for the client $c$. The number of useful packets from node $u$ that reach the client $c$ in this case is given by the rate $N_c(u) = b_o(u) \cdot (1-\epsilon_c(u)) $. When this rate is uniform, the decoding delay when the node $u$ is the only source of information is given by:

\begin{equation} \displaystyle
        t_c (u) = \frac{G}{b_o(u) \cdot (1-\epsilon_c(u)) }
\label{eq:t_c1}
\end{equation} 

Finally, the average decoding delay at client $c$ is computed by considering all the sources and NC nodes as independent sources of information with uniform useful rates $1/t_c (u)$. We can write the decoding delay as 
 
\begin{equation} \displaystyle
        t_c = \frac{1}{\displaystyle \sum_{u \in S}{\frac{1}{t_c (u)}}} ,
\label{eq:t_composite}
\end{equation}
where $S$ is the set of sources and NC nodes. 

\floatname{algorithm}{Algorithm}
\begin{algorithm}
\caption{Delay computation algorithm}
\label{algo:algo_innovative_refinedrates}
\begin{algorithmic}[1]
  \STATE Initialize replication rates for every node: $\hat{R}(u) = \frac{b_o(n)}{b_i(n)}$.
	\REPEAT
    \FOR{each client node $c$}
        \STATE Compute $\epsilon_c(u)$ for sending nodes $u$ from Eq. (\ref{eq:p0}).
        \STATE Compute $N_c(s) = b_o(s) \cdot (1- \epsilon_c(s))$ for all sources $s$.        
        \FOR{each NC node $u$}
                \STATE Compute $t_c(u)$ using Eqs. (\ref{eq:Pr<N})-(\ref{eq:tbar}) setting node $u$ in SF mode.
                \STATE Compute $t_c(u)^{'}$ using Eqs. (\ref{eq:Pr<N})-(\ref{eq:tbar}). setting node $u$ in silent mode.
                \STATE Compute $\Delta t_c(u)= t_c(u)-t_c(u)^{'}$.
                \STATE Compute $\Delta N_c(u)=1/\Delta t_c(u)$.
                \STATE Compute $N_c(u)$ using Eq. (\ref{eq:D_cn}).
                \IF{$N_c(u) < b_o(u)$}
                        \STATE Compute $t_c(u)$ using Eqs. (\ref{eq:Pr<N})-(\ref{eq:tbar}) setting node $u$ in NC mode.
                       \ELSE
                        \STATE Compute the expected decoding delay $t_c (u)$ using Eq. (\ref{eq:t_c1}).
                \ENDIF
        \ENDFOR
        \STATE Compute the average decoding delay $t_c$ considering all sources and NC nodes simultaneously, with Eq. (\ref{eq:t_composite}).
      \ENDFOR
      
      \FOR{each SF node $u$}                   
	\STATE Estimate the total number of packets received by each node, per generation: $\displaystyle |\mathcal{N}(u)| = b_i(u) \cdot \max_c{t_c}$.
	\STATE Update the replication rate $\hat{R}(u)$ with Eq. (\ref{eq:Rech}).
      \ENDFOR
	\UNTIL {Until convergence of $t_c$.}
\end{algorithmic}
\end{algorithm}

\section{Selective placement of NC nodes}
\label{sec:algos}

Equipped with methods to estimate the decoding delay on the overlay network, we can design algorithms to decide which nodes in the network should perform network coding. We address the problem of placing $A$ network coding nodes in the overlay network, such that the average delay observed by the clients is minimized. This is typically achieved by selecting network coding nodes such that the packet replication rate is decreased and the innovative flow rate in the network is increased.  
 
However, the optimal selection of the NC nodes is known to be an NP-hard problem \cite{langberg2009computational}. Hence, we design a greedy approach that iteratively searches for the optimal placement of a new network coding node while all the previously added NC nodes are fixed. The candidate nodes for implementing network coding are all the remaining SF nodes in the overlay network. Our node selection algorithm examines all the SF nodes backwards from the clients to the servers. It selects the SF node whose transformation into an NC node brings the highest benefit for the clients. This procedure is repeated until all the $A$ NC nodes have been selected. 

We propose now two variants of the iterative selection algorithm that both use Algorithm \ref{algo:algo_innovative_refinedrates} for computing the innovative flow rate, but differ in their view of the network resources. The first algorithm assumes that a central node possesses a global knowledge of the network; it iteratively selects the network coding nodes in a centralized manner. When global network knowledge is not a reasonable assumption, the centralized algorithm still serves as a performance benchmark for other greedy NC node placement algorithms. The second algorithm uses only a local view of the network resources at each node for computing the gains in innovative rate and decoding delay. This algorithm is probably more realistic in practice and can be implemented in a distributed way. 

%\subsection{Global information}
\label{sec:algo_centr}

In more details, the centralized algorithm uses the knowledge about the full network and available resources in order to determine the number of innovative packets received by each client. It leads to the iterative selection of $A$ NC nodes by computing at each stage the benefit of turning any of the SF nodes into an NC node. The candidate node that brings the highest innovative flow rate with its transformation is selected as a new NC node. The algorithm is described in Algorithm \ref{algo:algo_full}.

\floatname{algorithm}{Algorithm}
\begin{algorithm}
\caption{Centralized NC node selection}
\label{algo:algo_full}
\begin{algorithmic}[1]
    \FOR{i = 1 to A}
        \FOR{each node $u$ in the set of SF nodes.}
            \STATE Turn temporarily $u$ into a NC node
            \STATE Estimate the average decoding delay at the clients $t_c$ (using Algorithm \ref{algo:algo_innovative_refinedrates}).
            \STATE Turn $u$ back into a SF node.
        \ENDFOR
        \STATE Select the node $u^{\ast}$ that minimizes the decoding delay, i.e., $u^{\ast} = \argmin_u \sum_c t_c$
        \STATE Turn permanently $u^{\ast}$ into a NC node.
    \ENDFOR
\end{algorithmic}
\end{algorithm}

%\subsection{Local information}
\label{sec:algo_distrib}

The second algorithm relaxes the assumption that a central node is aware of the full network status. Instead, the nodes only use local network information for the estimation of the innovative flow rate. We define a neighborhood around each node. Then, an algorithm similar to the centralized solution above is applied in each neighborhood in order to determine the benefits of turning SF nodes into NC nodes. In particular, each node uses the estimation of the reception probability $\epsilon_c(u)$ that is given by all nodes $u$ in the neighborhood and computes an estimation of the decoding delay based on local information, \textit{i.e.}, the capacities and the loss rates of the subnetwork around the node $u$. Note that $\epsilon_c(u)$ is also calculated  considering only the statistics of node's $u$ neighborhood.
These estimations are transmitted periodically to a central agent, which finally makes the decision on the placement of new NC nodes. The procedure is summarized in Algorithm \ref{algo:distributed}.

\floatname{algorithm}{Algorithm}
\begin{algorithm}
\caption{NC node selection with local information} \label{algo:distributed}
\begin{algorithmic}[1]
    \FOR{i = 1 to A}
        \FOR {every SF node $u$}
            %\STATE Estimate from local information the average delay $t_c (n)$ at the clients.
            \STATE Temporarily transform node $u$ into an NC node.
            \STATE Estimate the average delay $\hat{t}_c (u)$ at the clients using Algorithm \ref{algo:algo_innovative_refinedrates} with local information.
            %\STATE Compute the benefit of the transformation: \\Ê$\mathcal{B}_c(n) = \hat{t_c}(n)- t_c (n)$
            \STATE Transform node $u$ back into an SF node.
            %\STATE Transmit $\mathcal{B}_c(n)$ to a central agent.
            \STATE Transmit $\hat{t}_c(u)$ to a central agent.
        \ENDFOR
        \STATE Select the node $u^{\ast}$ that maximizes the innovative rate, $u^{\ast} = \argmax_u \sum_c \hat{t}_c(u)$
        \STATE Turn permanently $u^{\ast}$ into a NC node.
    \ENDFOR
\end{algorithmic}
\end{algorithm}
%\vspace{-0.3cm}

%*****************************************************************************************************
%*****************************************************************************************************
%\subsection{Discussion}
%*****************************************************************************************************
%*****************************************************************************************************

Both algorithms permit to select a few network coding nodes in the system, such that the coding delay and overall computational complexity in the network is limited. At the same time, the system maintains a high innovative rate for sustained streaming performance. The choice of the number of NC nodes is typically determined by the admissible delay or tolerable complexity in the network. For example, constraints on decoding delay impose a limit on the maximum number of NC nodes in the system. However, the problem of determining the optimal number of NC nodes is out of the scope of this paper. We rather assume that the number of coding nodes or helpers in the streaming system is given a priori. The proposed algorithms then solve the problem of placing efficiently the NC nodes in the overlay network. 

Finally, it has to be noted that the second algorithm is not fully distributed, as it still uses a central agent to select the NC nodes. However, since it uses only local information, the proposed solution is certainly amenable to a fully distributed algorithm. One could imagine that each node decides independently if it should implement network coding or not, by comparing the local estimation of the gain in innovative rate to a pre-defined threshold. Alternatively, a distributed consensus solution could be deployed for a coordinated selection of the NC nodes with minimal information exchange between the overlay nodes. 

\section{Simulation results}
\label{sec:res}

\subsection{Setup}

In this section, we analyze the performance of the proposed NC node selection algorithms for the transmission of video streams in overlay networks. We generate overlay networks based on realistic network bandwidth values and adjacency measurements from the Planet Lab \cite{PlanetLab}, as provided in a snapshot of their network taken on 24 Nov. 2009 by their Scalable Sensing Service ($S^3$) \cite{S3}. The networks under consideration have one source node, three client nodes and a variable number of intermediate nodes. We create network topologies in the following way. First, the source nodes are positioned, then the nodes are randomly added one-by-one to the topology. For every new node, four nodes are randomly chosen as parent nodes. However, if the new node is not directly connected to any of the selected parents according to the Planet Lab measurement data, the node is removed and a new node is selected. After all nodes have been added to the network, the nodes that cannot be reached by the source and the nodes that are not connected with any client are removed. The resulting network graphs are directed acyclic by construction. The edge capacity of each link is set to $1/200$ of the Planet Lab capacity values, in order to get realistic values for the link bandwidth. Finally, the packet loss rate of each link is set to $5\%$. 

The network coding operations are performed in a Galois field of size GF$(2^{8})$ since this field size has been shown to result in a good compromise between performance (packets are innovative with high probability) and information overhead \cite{PracticalNC03}. The generation size is set to 32 packets, which is reasonable for real time video streaming applications.  The packet size is 512 bytes. The decoding is performed by gaussian elimination. Since we are interested in analyzing the performance in terms of decoding delay, we compute the average delay as the time needed by each client to receive 32 linearly independent packets (\textit{i.e.}, a complete generation). This is the minimal number of packets for the clients to decode the source information. We use a set of 10 random networks that consist of 32 to 56 nodes with two different radii of 6 and 8 hops (i.e., the maximum distance between any pair of nodes in the network is 6 or 8 hops). We consider the placement of up to 10 NC nodes in each network. In each case, the performance results are averaged over 100 simulations performed using the NS-3 \cite{NS3} network simulator. 

We evaluate the performance of the centralized and semi-distributed algorithms (resp. the Algorithm \ref{algo:algo_full} and Algorithm \ref{algo:distributed}) denoted as ``ALGO2'' and ``ALGO3'' respectively). We compare them with a greedy search algorithm (ALGO2R) similar to Algorithm \ref{algo:algo_full} but that uses the actual delays obtained by NS-3 simulations instead of the delay estimates from Algorithm \ref{algo:algo_innovative_refinedrates} for the node selection. We also compare to a baseline scheme called in the following as ``RANDSEL'' that randomly places the NC nodes in the network. For the sake of completeness, we finally study the performance of a scheme where all nodes are NC nodes (this scheme is called ``All nodes NC''), a Linear NC scheme \cite{PracticalNC03} and a scheme with only SF nodes. The performance of the later is equal to the theoretical maximum that can be sustained when only routing is enabled and can be found by typical maximum flow algorithms). 

Note that the minimum delay and maximum effective throughput obtained with Linear NC are computed by considering that a high capacity hyper-source node connected with all sources. In this case, the overall throughput is computed as the sum of the throughputs from the hyper-source to each client node. For the routing case, we consider as well a hyper-sink node linked with all client nodes, and we compute the maximum throughput between the hyper-source and hyper-sink with standard graph maximum flow algorithms. We should point out that the links connecting the hyper nodes with the networks sources and clients are error free and have infinite capacities, so they do not introduce extra delays. 

\subsection{Decoding delay}

\begin{figure*}[t]
\begin{center}
\begin{tabular}{cc}
~\includegraphics[width=0.4\textwidth]{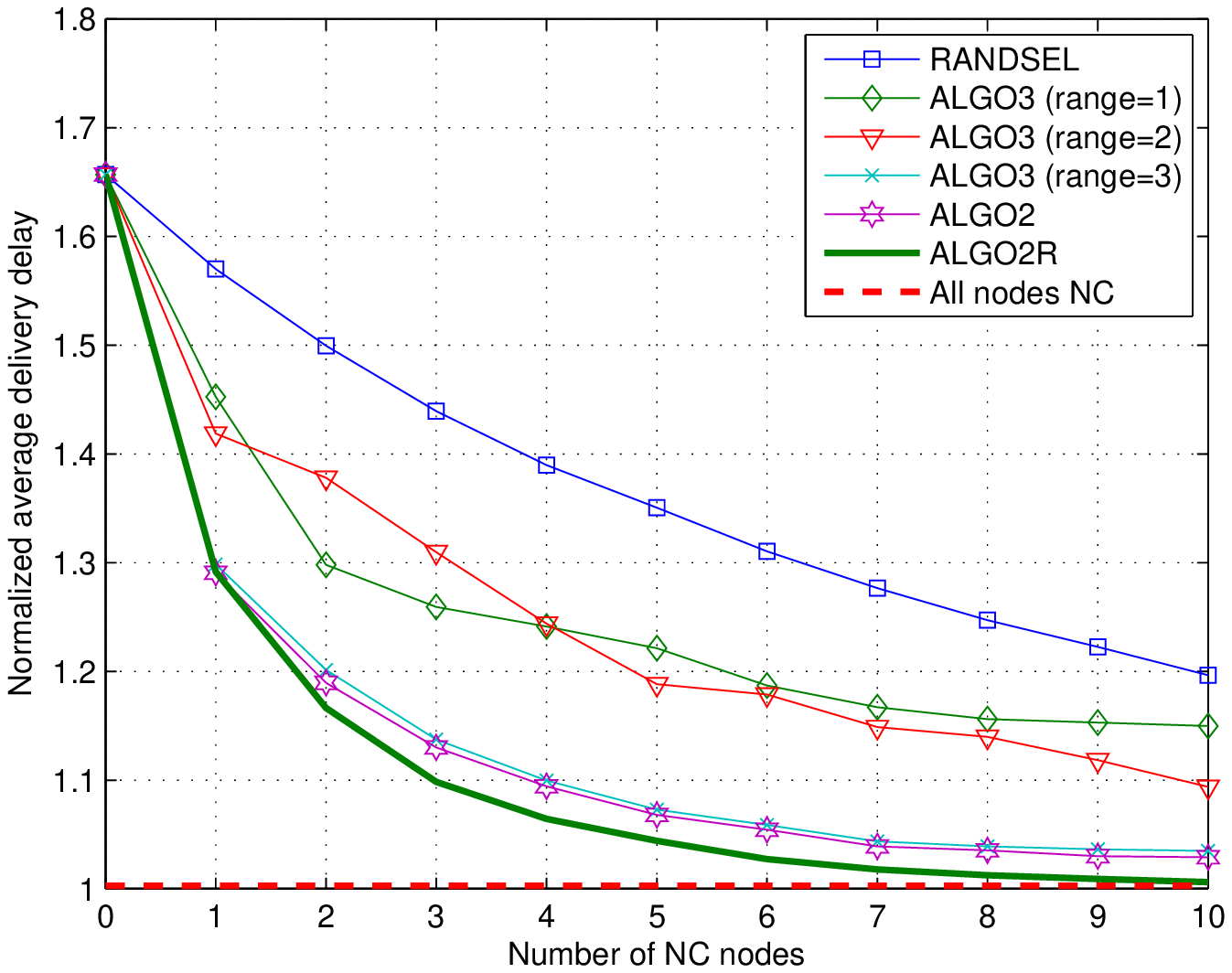}
~ & ~\includegraphics[width=0.4\textwidth]{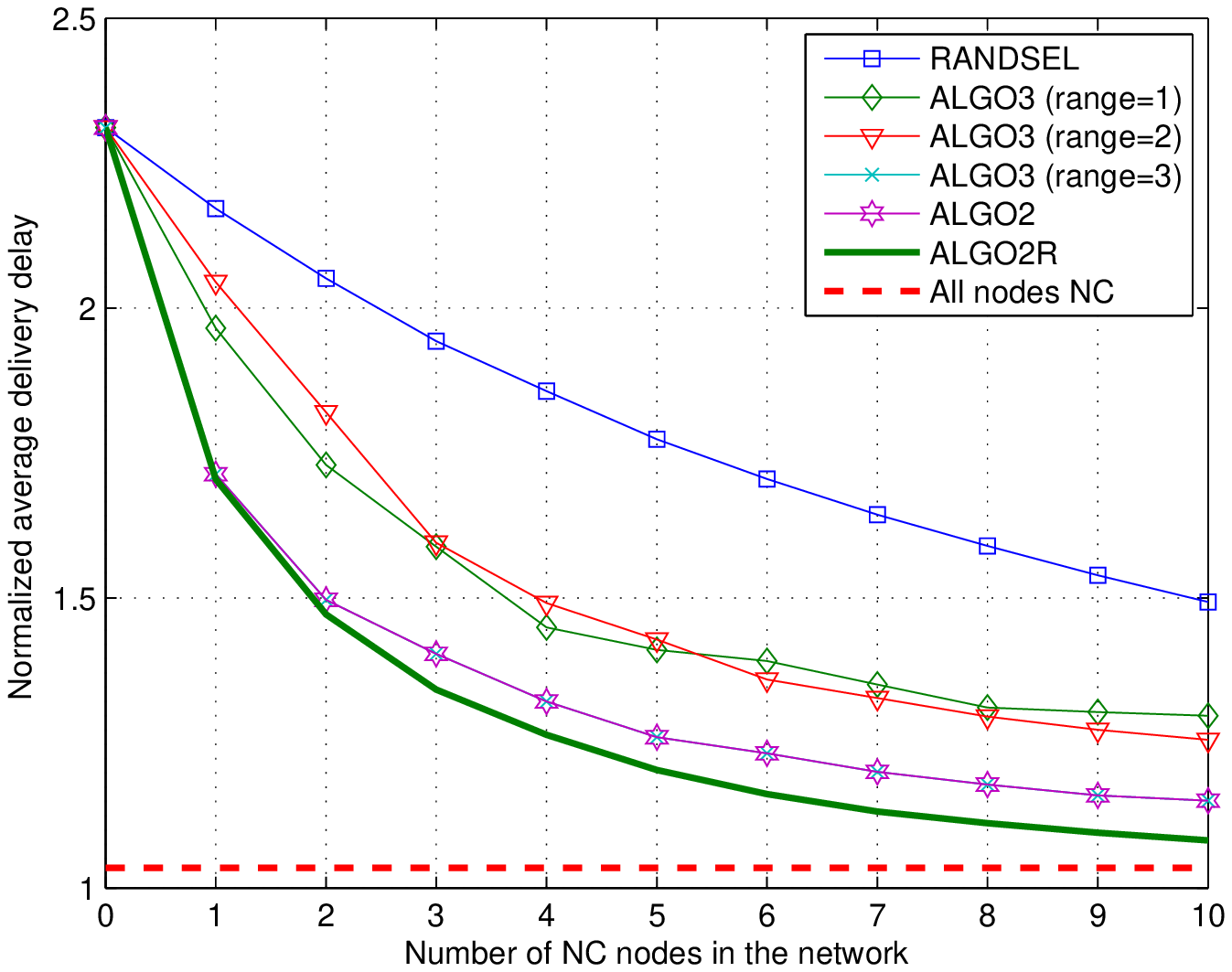}~\\
~(a)~&~(b)~\\
\end{tabular}
\end{center}
\caption{Normalized average decoding delays versus the number of NC nodes, in networks with maximum distance between any pair of nodes equal to: (a) six and (b) eight.}  
\label{fig:res_delays}
\end{figure*}

We first study the decoding delay for each of the algorithms. Figs. \ref{fig:res_delays} (a) and (b) illustrate the normalized average decoding delays for the network clients as a function of the number of NC nodes added in the network, for two different network sizes. The decoding delays are normalized to the performance obtained when all nodes perform network coding. We show the performance of Algorithm \ref{algo:distributed} for different sizes of the neighborhood in the local gain estimations. We observe a sharp reduction of the delivery times with the addition of the first few NC nodes for all the algorithms, but especially for the proposed algorithms. The gains become less important after a few NC nodes have been placed in the network. We can thus see that the NC nodes are well positioned in order to improve the delivery performance. The results also highlight the inefficiency of the RANDSEL algorithm, which becomes competitive with the other methods only for large number of NC nodes. Finally, we can see that the Algorithm \ref{algo:algo_full} performs similar to the ALGO2R. This confirms that the proposed delay estimation strategy in ALGO2 is accurate as it comes close to the actual delay values measured by the network simulator. 

\subsection{Innovative rate}

\begin{figure*}[t]
\begin{center}
\begin{tabular}{cc}
~\includegraphics[width=0.4\textwidth]{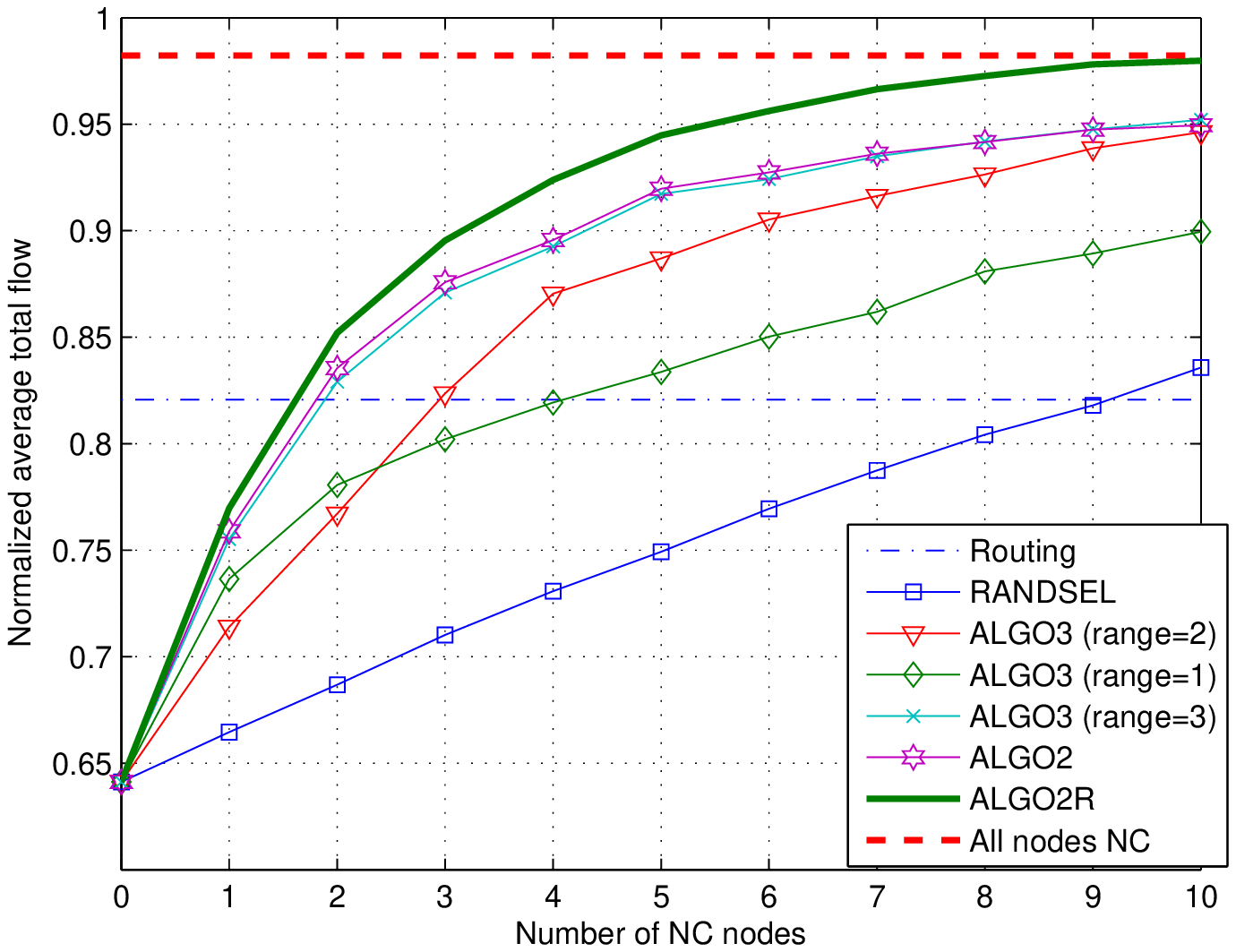}
~ & ~\includegraphics[width=0.4\textwidth]{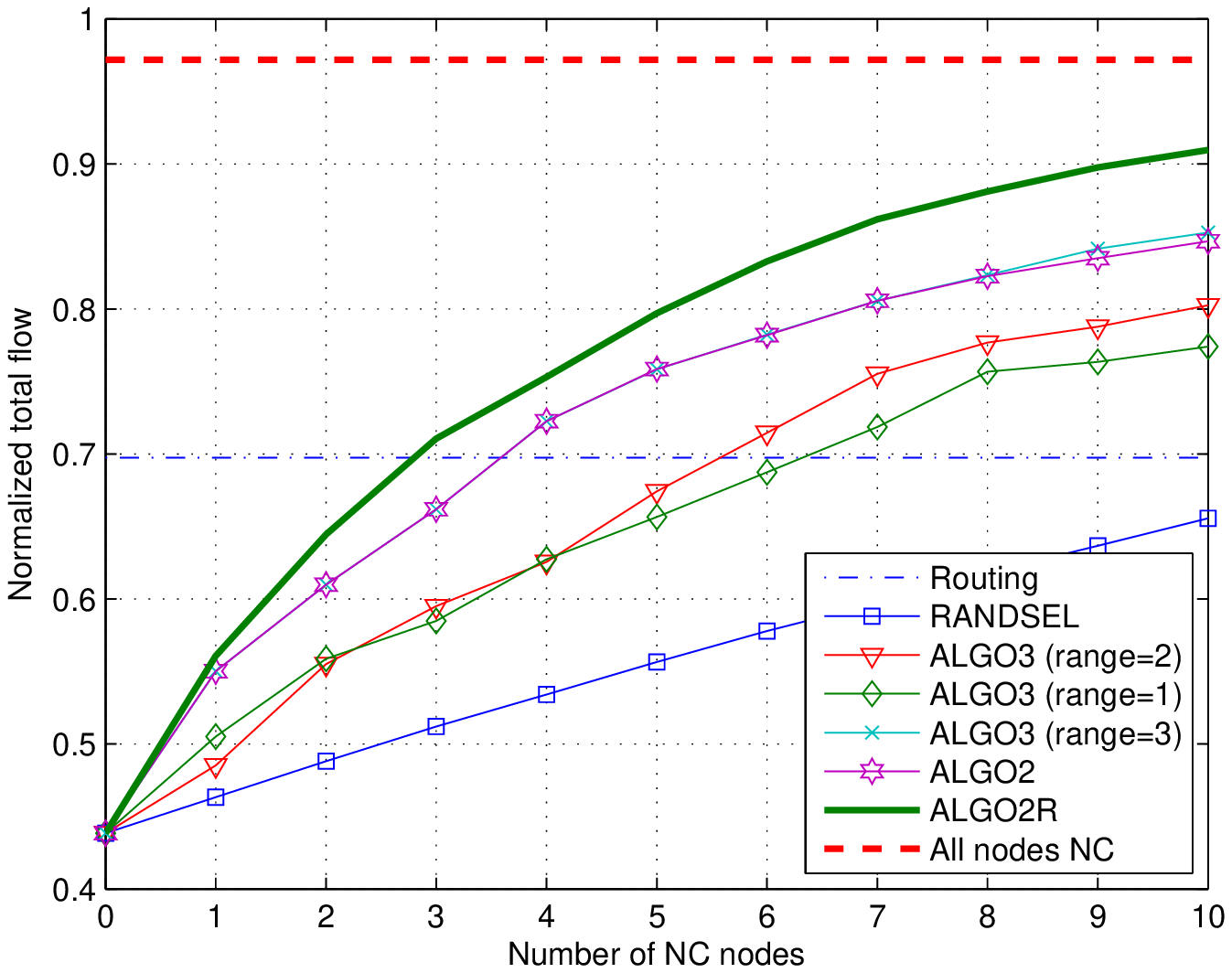}~\\
~(a)~&~(b)~\\
\end{tabular}
\end{center}
\caption{Normalized achievable throughput versus the number of NC nodes, in networks with maximum distance between any pair of nodes equal to: (a) six and (b) eight.}  
\label{fig:res_flows}
\end{figure*}

We further look at the average normalized effective throughput in the network (\textit{i.e.}, the number of useful packets received by the clients), as a function of the number of NC nodes in the network. Normalization is performed with respect to the effective throughput achieved by the scheme where all nodes perform network coding. Figs. \ref{fig:res_flows} (a) and (b) show the effective throughputs for two different network radii. The results confirm the earlier observations on the decoding delay performance. A few well selected nodes are able to bring a large throughput gain. Further performance improvements become less important as the number of NC nodes increase. We see also that the algorithms proposed in this paper provide the best performance among the schemes under comparison. The node selection algorithm with local information improves with the size of the neighborhood but generally stays close to the centralized algorithm when the neighborhood is sufficiently large. In the case where the neighborhood is limited to one node, the proposed algorithm still outperforms RANDSEL since the decisions are not totally blind. The reason for the inferior performance of the semi-distributed scheme compared to the centralized one simply comes from the fact that the local network statistics are not sufficient for accurately estimating the delay when the neighborhood is small. In addition, we can observe that a few NC nodes are sufficient for obtaining higher throughputs than the one in routing algorithms where the nodes simply forward packets randomly to their descendants. It confirms the fact that our methods are appropriate for deployment in low-cost networks, where a few helpers or network coding nodes are sufficient for improved throughput and efficient data delivery. Finally, we can observe that our algorithms tend to perform better in larger networks (i.e., larger radius values), as they are able to exploit more efficiently the available resources and the diversity in the overlay network. 

\subsection{Video quality}

\begin{figure*}[t]
\begin{center}
\begin{tabular}{cc}
~\includegraphics[width=0.4\textwidth]{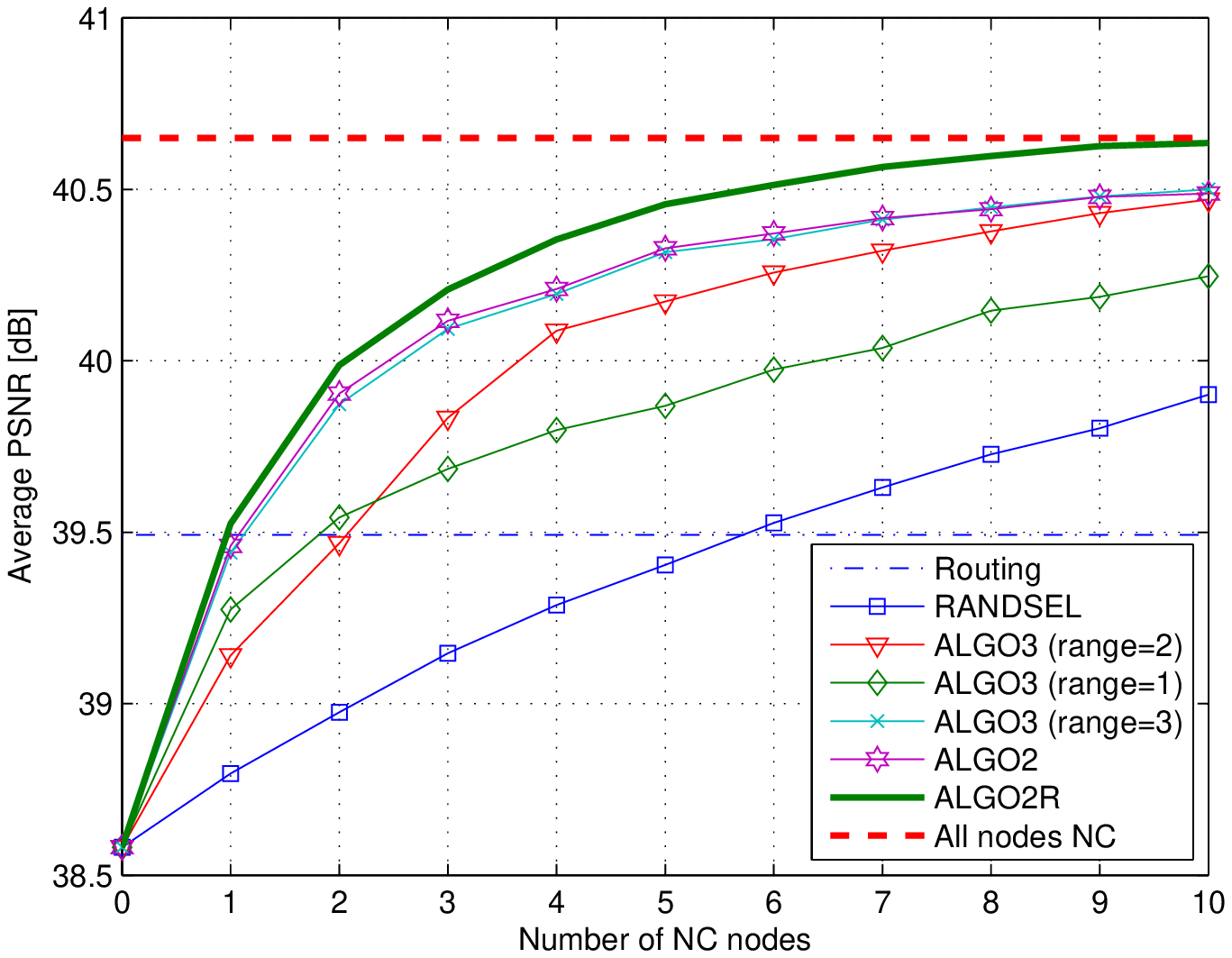}
~ & ~\includegraphics[width=0.4\textwidth]{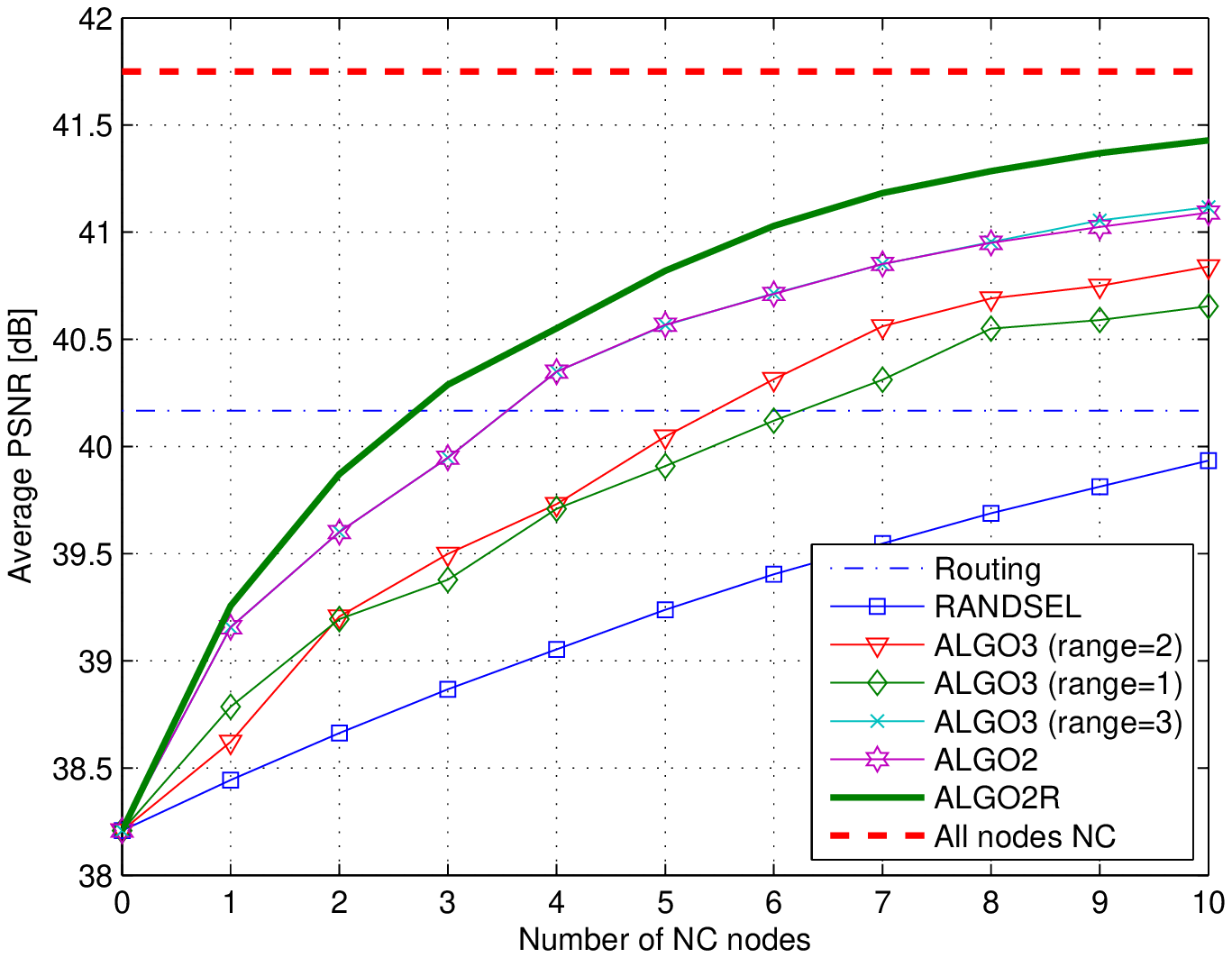}~\\
~(a)~&~(b)~\\
\end{tabular}
\end{center}
\caption{Average PSNR quality at clients versus the number of NC nodes, in networks with maximum distance between any pair of nodes equal to: (a) six and (b) eight.}
\label{fig:res_PSNR}
\end{figure*}

Finally, we study the performance of the video delivery schemes from the viewpoint of video quality. We estimate the average PSNR quality measured at the clients with respect to the number of NC nodes in the network for all methods under comparison in the transmission of the Foreman CIF sequence encoded by the the JM12.2 \cite{JVT122} of the H.264/AVC standard \cite{H264Standard}.  The quality is estimated by setting the encoding rate to the value of the network throughput in the different schemes.  The corresponding results are illustrated in Figs. \ref{fig:res_PSNR} (a) and (b) for networks with a radius of six and eight hops, respectively. It is interesting that the improved throughput values translate into higher PSNR quality which confirms the above observations about the benefits of proper NC node selection. We can have gains that exceed 1.5 dB with only two NC nodes, whereas we reach gains of 3 dB for seven NC nodes. The PSNR gains saturate as the number of NC nodes increases, but quality gains can still be noticed. As expected, larger gains are observed for the centralized algorithm, however, the semi-distributed algorithm offers significant gains as well. For a neighborhood of three hops, the performance of the semi-distributed even becomes identical to that of the centralized scheme. Finally, we see that RANDSEL gives small PSNR gains for a few NC nodes, which confirms the poor performance of a random selection of the network coding nodes and supports the development of effective selection algorithms.

\section{Related work}
\label{sec:related}

The problem of finding a minimal set of network coding points in a network has been mostly studied from a theoretical perspective so far. First, the special case of two source messages is examined in \cite{TavoryECCC03} where it is proved that the number of coding nodes is independent of the total number of network nodes. In \cite{FragouliColoring04}, the minimum number of network coding nodes is computed through graph coloring techniques. It is then shown in \cite{FragouliTIT2006} that the number of coding nodes is upper bounded by the number of receivers. A unification of network coding and tree packing theorems is further presented in \cite{WuTIT2006}, where network coding is restricted to pre-selected edges. These include only input edges of relay nodes and not the input edges of clients where simple routing is applied. This choice is made in order to achieve the min-cut max-flow limit of the network and save both processing and implementation complexity. The relation between links capacities and the number of coding nodes is investigated in \cite{CannonsTIT08}, where it is shown that in directed acyclic networks arbitrary amounts of gain can be noticed when subsets of nodes of arbitrary size are used for coding. Finally, the problem of finding network codes with a minimal number of encoding nodes has recently been studied in \cite{langberg2009computational} where the optimization problem is however shown to be NP-hard. 

While the previous work mostly consider that the network is fully known at a central node, a decentralized algorithm for minimizing the number of network coding packets flowing in a network has been presented in \cite{BhattadISIT05}. It also addresses the design capacity approaching network codes that minimize the set of network coding nodes. However, this algorithm does not provide any guarantee that the minimum set of network coding nodes can always be determined. While \cite{BhattadISIT05} consider capacity approaching codes without delay constraints, we rather use well performing network codes and consider the available resources in the network in order to select a set of network coding nodes, such that the overall delay is kept small in multimedia applications. The choice of  randomized network codes is mostly geared towards the implementation of practical distributed systems where large benefits are expected by the proper choice of a limited number of network coding nodes. 

In general, the previous works about the selection of coding nodes do not consider delay issues, which are most important in streaming applications. The problem of the selection of network processing nodes in multimedia streaming applications has been addressed in \cite{MingquanWuJSAP05} in a framework that is however slightly different than ours. The placement of a limited number of network-embedded FEC nodes (NEF) is considered in networks that are organized into multicast trees. The placement is chosen in order to enhance the robustness to transmission errors and to improve the network's throughput. NEF nodes first decode and successively re-encode the recovered packets in order to increase the symbol diversity. A greedy algorithm is proposed for placing NEF nodes. Although the proposed method is efficient, it is computationally expensive and unrealistic to be deployed in dynamic networks. In contrast to \cite{MingquanWuJSAP05}, we consider the placement of processing nodes in the more general case of overlay mesh networks with randomized network coding for distributed packet delivery.

Finally, game theoretic concepts are adopted in a recent work \cite{ThomosICASSP10} for developing socially optimal distributed algorithms that decide on the nodes that should combine packets. Specifically, incentives such as extra download bandwidth are given to network nodes in order to change their status to network coding and indirectly minimize the delays in the system. However, this algorithm does not offer any guarantee that limited resources will be used efficiently, since all the nodes may potentially desire to become network coding nodes. It is not appropriate when a certain number of network coding nodes shall be placed effectively in a network.

\section{Conclusions}
\label{sec:conclusions}

We have considered the problem of the placement of a predefined number of network coding nodes in a overlay media streaming system. We have proposed novel algorithms that iteratively select the best nodes for network coding such that the delay is decreased. The deployment of network coding gets positioned as a valid solution for exploiting the network diversity in streaming applications. We show that the selection of a small number of network coding nodes is able to provide an effective tradeoff between packet duplicates, decoding delay and computational complexity. The experimental evaluation in irregular and realistic networks shows that the proposed node selection schemes achieves the same throughput as a system where all nodes perform network coding, but with a dramatically smaller number of network coding nodes and hence lower complexity. In addition, we show that the quality of experience in video streaming applications is greatly improved by the proper selection of network coding nodes. Finally, the proposed algorithm is amenable to the implementation of distributed solutions that are able to adapt to the local characteristics of a dynamic network topology. It could also offer important insights in the design of effective media delivery solutions, where helpers nodes could be positioned in an overlay networks for maximizing the quality of service offered to the media clients.

\balance

\end{document}